\documentclass[12pt,a4paper]{article}
\usepackage{amsmath,amssymb,amsfonts,amsthm,bm,bbm,cancel,wasysym}
\usepackage{epsfig,graphics,graphicx,epstopdf}
\usepackage{array,booktabs,colortbl,colordvi,multirow}
\usepackage{colordvi,color,xcolor}
\usepackage{rotating}

\newcommand{\be}{\begin{equation}} 
\newcommand{\ee}{\end{equation}} 
\newcommand{\bea}{\begin{eqnarray}}  
\newcommand{\eea}{\end{eqnarray}}
\newcommand{\nn}{\nonumber \\} 

\newcommand{\lrD}{~\!\overset{\leftrightarrow}{\hspace{-0.1cm}D}\!}
\newcommand{\lD}{~\!\overset{\leftarrow}{\hspace{-0.1cm}D}\!}
\newcommand{\rD}{~\!\overset{\rightarrow}{\hspace{-0.1cm}D}\!}
\newcommand{\lrDa}{~\!\overset{\leftrightarrow}{\hspace{-0.1cm}D}\!^{\!~a}}

\newcommand{\refeq}[1]{Eq.~\!(\ref{#1})}

\newcommand{\units}[1]{~\mathrm{#1}}

\newcommand{\bfmath}[1]{{\mbox{\boldmath{$#1$}}}}

\newcommand{\ctoprule}{\toprule[0.5mm]}
\newcommand{\cbottomrule}{\bottomrule[0.5mm]}
\newcommand{\cmrule}{\midrule[0.25mm]}
\newcommand{\crowcolor}{\rowcolor[rgb]{0.9,0.9,0.9}}
\newcommand{\graycell}{\cellcolor[rgb]{0.9,0.9,0.9}}

\newcommand{\mt}[1]{\mathrm{#1}}

\begin{document}


\begin{flushright}
DESY 15-110\\
\today
\end{flushright}
\vspace*{2mm}

\renewcommand{\thefootnote}{\fnsymbol{footnote}}
\setcounter{footnote}{1}

\begin{center}

{\Large {\bf Renormalization Group Constraints on New Top\\
Interactions from Electroweak Precision Data}}\\
\vspace*{0.5cm}
{\bf J.\ de Blas${}^a$\footnote{E-mail: Jorge.DeBlasMateo@roma1.infn.it},}
{\bf M.\ Chala${}^b$\footnote{E-mail: mikael.chala@desy.de}} and 
{\bf J.\ Santiago${}^{c}$\footnote{E-mail: jsantiago@ugr.es}}

\vspace{0.25cm}

{\it ${}^a$INFN, Sezione di Roma, Piazzale A. Moro 2, I-00185 Rome, Italy}\\
\vspace{0.5cm}
{\it ${}^b$DESY, Notkestrasse 85, 22607 Hamburg, Germany}\\
\vspace{0.5cm}
{\it ${}^c$Departamento de F\'{\i}sica Te\'orica y del Cosmos and CAFPE,\\
Universidad de Granada, Campus de Fuentenueva, E-18071 Granada, Spain}\\
\vspace{0.5cm}

\end{center}
\vspace{.05cm}

\begin{abstract}
 
\noindent  Anomalous interactions involving the top quark contribute to some of the most
difficult observables to directly access experimentally. They can give
however a sizeable correction to very precisely measured observables 
at the loop level. Using a model-independent effective
Lagrangian approach, we present the leading indirect constraints on 
dimension-six effective operators involving the top quark from
electroweak precision data. They represent the most stringent
constraints on these interactions, some of which may be directly testable in
future colliders.

\end{abstract}

\renewcommand{\thefootnote}{\arabic{footnote}}
\setcounter{footnote}{0}


\section{Introduction}
\label{section_Intro}

Once the presence of the Higgs boson has been firmly established at the
Large Hadron Collider (LHC) the focus has turned towards the discovery
of new physics (NP). Already at Run 1 the absence of significant
deviations from the Standard Model (SM) predictions has
put stringent bounds on the mass scale of NP. 
Indirect constraints on new particles, when they are beyond the
kinematic reach of the LHC, are becoming competitive and in some
cases complementary to the indirect constraints from electroweak
precision data (EWPD)~\cite{deBlas:2013qqa}. It is expected that, with
the increased energy available during Run 2, any new particle within the
kinematic reach will be discovered, or very stringent constraints will
be placed in case they cannot be directly produced. Still,
there are certain interactions that, even with the increased energy, 
will be very difficult to directly probe at the LHC with a significant
precision. A notable example is that of the interactions involving the top quark
that we consider in this work.

Many well-motivated models of NP predict the largest deviations from
the SM results in processes involving the top quark. Using an
effective Lagrangian description, several groups have
studied the potential of the LHC to constrain higher-dimensional
operators involving the top quark in single and pair top
production~\cite{eff:top},
including in some cases next-to-leading order
predictions (see~\cite{eff:top:nlo} and references therein). 
Some of these interactions can be directly constrained to
a reasonable accuracy for the first time at the LHC. However, the
complexity of the $t(\bar{t})$ system limits the precision that one
can achieve with these direct probes. In some cases, certain
higher-dimensional operators are just inaccessible at the LHC. Many of
these interactions, however, contribute at the loop level to EWPD. The
very stringent constraints that can be derived from EWPD, together
with the fact that the relevant coupling is usually the top Yukawa
coupling, can compensate for the loop suppression, thus producing the
most stringent constraints on many higher-dimensional operators involving
the top quark. In particular, this includes those that cannot be directly probed
at the LHC. 

In this article we use EWPD to place bounds on
NP interactions. We use a model-independent effective
Lagrangian approach going beyond the usual analysis of dimension-six
operators correcting precision observables at the tree level. Analyses including 
one-loop contributions from higher-dimensional operators have been done for a
small subset of purely bosonic
operators in the past \cite{De Rujula:1991se}. 
Here we use the calculation of the
renormalization group equations (RGE) for the entire dimension-six 
effective
Lagrangian~\cite{Jenkins:2013zja,Jenkins:2013wua,Alonso:2013hga} (see
also~\cite{Grojean:2013kd}) to
determine, without restricting to any particular set of
operators, which interactions can be constrained by EWPD at the
$O(0.1)$ or better. 
This precision can be achieved by higher-dimensional operators that
give a sizeable loop 
contribution to
EWPD. If we further neglect operators that can be
directly probed in current or past experiments, we are basically left with
dimension-six operators involving top quarks.
We will show that
indirect constraints from EWPD can be quite stringent for these
interactions, some of which could be tested in future lepton colliders.

We discuss the global fit to EWPD, including the leading, currently
unconstrained, loop effects in section~\ref{global_fig:sec}. The
corresponding constraints on the coefficients of the dimension-six
operators involving the top quark are presented in
section~\ref{constraints:sec} and we conclude in
section~\ref{conclusions:sec}.


\section{The global electroweak fit for new physics to dimension six}
\label{global_fig:sec}

Assuming that NP is heavier than the
energies currently probed by experiments,
its effects can be described by an effective Lagrangian,
\be
{\cal L}_\mt{Eff}={\cal L}_\mt{SM}
+\frac{1}{\Lambda^2} {\cal L}_6+\dots,
\label{EffL}
\ee
where ${\cal L}_d=\sum c_i {\cal O}_i$, with ${\cal O}_i$ Lorentz and
SM gauge invariant operators of mass dimension $d$ built from the SM
fields. $\Lambda$ stands for the cut-off scale of the effective theory, and we
have neglected lepton number violating effects. 
There is
a total of 59 operators (up to flavor indices) at dimension six~\cite{Grzadkowski:2010es}.
However,  
only a few of these directly contribute at leading order
to EWPD. We consider in our analysis the following set of EWPD:  
\be
\begin{split}
\left\{M_H,~\!m_t,~\!\alpha_S(M_Z^2),~\!\Delta \alpha_\mt{had}^{(5)}(M_Z^2),~\!M_W,~\!\Gamma_W,~\!\mt{Br}_{W}^{e\nu,\mu\nu,\tau\nu},~\!M_Z,~\!\Gamma_Z,~\sigma_\mt{had},~\!R_{e,\mu,\tau},\right.\\
\left.A_{FB}^{e,\mu,\tau},~\!A_{e,\mu,\tau} \mt{(SLD)},~\!A_{e,\tau}\mt{(}P_\tau\mt{)},~\!R_{b,c},~\!A_\mt{FB}^{b,c},~\!A_{b,c},~\!A_\mt{FB}^s,~\!A_s,~\!R_u/R_{u+d+s},~\!Q_\mt{FB}^\mt{had}\right\},
\label{EWPO}
\end{split}
\ee
whose definition, experimental values, errors and correlation matrices
are taken from \cite{Aad:2015zhl}. 

With these observables we perform a fit to the SM extended
with the relevant set of dimension-six interactions that we will introduce below. 
The SM predictions for EWPD are computed including the latest
theoretical developments~\cite{Awramik:2003rn}. These are then corrected by the NP dimension-six
operators, and compared to the experimental measurements via the usual $\chi^2$ function,
which we minimize and use to determine the bounds on the new interactions.
In our fits both the NP and the SM parameters are allowed to float.
In order to
include dimension-six effects in a consistent way, we limit the
contributions to EWPD to order $1/\Lambda^2$. These may come either
from the direct interference of the NP amplitudes with the SM ones in the
electroweak precision observables, or from NP contributions to the physical 
processes where some of the SM input parameters
are determined (e.g. $G_F$ as extracted from muon decay), which then
propagate to all observables. Effects of order 
$1/\Lambda^4$ would be comparable to interference effects coming from 
dimension-eight operators and have been therefore neglected. Given the 
resulting bounds, this seems in general a good approximation. An analysis
 of the error induced by neglecting
such contributions, along the lines explained in e.g. \cite{Berthier:2015oma}, 
goes beyond the scope
of this paper. 

The analysis of bounds on dimension-six interactions from EWPD has
been considered extensively in the
literature (see for instance~\cite{Han:2004az}). 
At the tree level,
only the following coefficients of 
dimension-six operators
modify the observables in (\ref{EWPO}): 
\be
c_\mt{EWPD}^\mt{tree}=\left\{c_{\phi l}^{(1)},~\!c_{\phi q}^{(1)},~\!c_{\phi e}^{(1)},~\!c_{\phi u}^{(1)},~\!c_{\phi d}^{(1)},~\!c_{\phi l}^{(3)},~\!c_{\phi q}^{(3)},~\!c_{\phi D},~\!c_{WB},~\!c_{ll}^{(1)}\right\}.
\label{EWPDdim6}
\ee
The definition of the operator basis used in this work follows closely the
original one in \cite{Grzadkowski:2010es} except for the
four-fermion sector for which we use the one 
in Appendix A of \cite{deBlas:2014mba}.\footnote{For completeness,
  the dimension-six operators in (\ref{EWPDdim6}) are presented here:  
\!${\cal O}_{\phi \psi}^{(1)}\!\!=\!\!(\phi^{\dagger}i\lrD_\mu \phi)(\overline{\psi} \gamma^\mu \psi)$, ${\cal O}_{\phi\psi}^{(3)}\!\!=\!\!(\phi^{\dagger} i\lrDa_\mu \phi)(\overline{\psi_L} \gamma^\mu\sigma_a \psi_L)$ (with $\lrD_\mu=\rD_\mu - \lD_\mu$ and $\lrDa_\mu=\sigma_a \rD_\mu - \lD_\mu \sigma_a$), ${\cal O}_{\phi D}\!\!=\!\!(\phi^{\dagger}D_\mu \phi)((D^\mu \phi)^{\dagger}\phi)$, ${\cal O}_{{WB}}\!\!=\!\!(\phi^\dagger\sigma_a \phi)~W^a_{\mu\nu}B^{\mu\nu}$ and ${\cal O}_{ll}^{(1)}\!\!=\!\!\frac 1 2 \left(\overline{ l_L} \gamma_\mu l_L\right)\left(\overline{l_L} \gamma^\mu l_L\right)$.
}
At the loop level, however, many other dimension-six operators
contribute to EWPD, 
including some to which we have
little or no direct experimental access. In this latter case, the high
precision of the EWPD can compensate the loop suppression and give the
most stringent constraints on these operators. The complete one-loop
calculation of the EWPD including dimension-six operators is beyond
the scope of this article but the logarithmically enhanced
contributions can be computed by means of the RGE recently computed
in~\cite{Jenkins:2013zja,Jenkins:2013wua,Alonso:2013hga}.  
An analysis of the loop-improved
electroweak constraints on the dimension-six interactions
in (\ref{EWPDdim6}) will be presented elsewhere
\cite{UpdateEWfitdim6}. Here we want to focus on those operators that,
as explained above, have not been directly probed by any experiment yet, and to
determine which ones can be constrained to a reasonable accuracy with
current data.

The largest RGE effects on EWPD are those proportional to the top
Yukawa coupling, $y_t$. 
(The strong coupling $g_3$ does not
enter in any of the anomalous dimensions for the interactions in (\ref{EWPDdim6})~\!.) 
Inspecting the RGE for the operators in (\ref{EWPDdim6}) and looking
for $y_t$ effects then allows us to identify which ones
could be
significantly constrained from the low-energy bounds. Restricting to
operators that have not been directly tested in experiments so far
leaves us with the following set, containing always the top 
quark:\footnote{Apart from these, we will also comment on 
another operator (see footnote 3): the pure scalar interaction
  ${\cal O}_{\phi \Box}=(\phi^\dagger \phi) \Box (\phi^\dagger
  \phi)$, which does not contribute to EWPD proportionally to $y_t$, but enters
  in the anomalous dimension of $c_{\phi D}$
  with a large coefficient $\sim 20 g_1^2/3$~\cite{Alonso:2013hga}. Although this operator can
  be in principle tested in Higgs physics, its constraints are still very 
  weak~\cite{deBlas:2014ula}.}
\bea
{\cal O}_{\phi q}^{(1)}\!\!\!\!&=&\!\!\!\!(\phi^\dagger i\lrD_\mu \phi)(\overline{q_L}\gamma^\mu q_L),
\qquad\,{\cal O}_{\phi q}^{(3)}\!=\!(\phi^\dagger i\lrDa_\mu \phi)(\overline{q_L}\gamma^\mu \sigma_a q_L),\nonumber\\[0.1cm]
\!\!{\cal O}_{\phi u}^{(1)}\!\!\!\!&=&\!\!\!\!(\phi^\dagger \lrD_\mu \phi)(\overline{u_R}\gamma^\mu u_R),\nonumber\\[0.1cm]
%
{\cal O}_{lq}^{(1)}\!\!\!\!&=&\!\!\!\!(\overline{l_L}\gamma_\mu l_L)(\overline{q_L}\gamma^\mu q_L),
\qquad\,\,\,\,{\cal O}_{lq}^{(3)}\!=\!(\overline{l_L}\gamma_\mu \sigma_a l_L)(\overline{q_L}\gamma^\mu \sigma_a q_L),\nonumber\\[0.1cm]
{\cal O}_{eu}\!\!\!&=&\!\!\!\!(\overline{e_R}\gamma_\mu e_R)(\overline{u_R}\gamma^\mu u_R),
\qquad\,\!\!{\cal O}_{lu}~\!=\!(\overline{l_L}\gamma_\mu l_L)(\overline{u_R}\gamma^\mu u_R),\nonumber\\[0.1cm]
{\cal O}_{qe}\!\!\!&=&\!\!\!\!(\overline{q_L}\gamma_\mu q_L)(\overline{e_R}\gamma^\mu e_R),\nonumber\\[0.1cm]
%
{\cal O}_{qq}^{(1)}\!\!\!\!&=&\!\!\!\!\frac 12(\overline{q_L}\gamma_\mu q_L)(\overline{q_L}\gamma^\mu q_L),
\qquad \!\!{\cal O}_{qq}^{(8)}\!=\!\frac 12(\overline{q_L}\gamma_\mu T_A q_L)(\overline{q_L}\gamma^\mu T_A q_L),\nonumber\\[0.1cm]
{\cal O}_{uu}^{(1)}\!\!\!\!&=&\!\!\!\!\frac 12(\overline{u_R}\gamma_\mu u_R)(\overline{u_R}\gamma^\mu u_R),
\quad \!\,\!{\cal O}_{ud}^{(1)}=\!(\overline{u_R}\gamma_\mu u_R)(\overline{d_R}\gamma^\mu d_R),\nonumber\\[0.1cm]
{\cal O}_{qu}^{(1)}\!\!\!\!&=&\!\!\!\!(\overline{q_L}\gamma_\mu q_L)(\overline{u_R}\gamma^\mu u_R),
\qquad{\cal O}_{qd}^{(1)}\!=\!(\overline{q_L}\gamma_\mu q_L)(\overline{d_R}\gamma^\mu d_R),
\nonumber\\[0.1cm]
{\cal O}_{uB}\!\!\!\!&=&\!\!\!\!(\overline{q_L}\sigma^{\mu\nu} u_R) \tilde{\phi} B_{\mu\nu},
\qquad\quad\,{\cal O}_{uW}\!=\!(\overline{q_L}\sigma^{\mu\nu} \sigma_a u_R) \tilde{\phi} W_{\mu\nu}^a.
\label{TopDim6}
\eea

An analysis of EWPD constraints on a small subset of
  these operators was performed in~\cite{Zhang:2012cd}. However, at one
  loop only non-logarithmic finite contributions were included and the
  corresponding bounds are much weaker.
Regarding the operators ${\cal O}_{\phi q}^{(1),(3)}$, we
will only consider the combination ${\cal O}_{\phi q}^{(t)}\equiv ({\cal O}_{\phi
  q}^{(1)}-{\cal O}_{\phi q}^{(3)})_{tt}$, which (up to corrections
suppressed by products of $V_{td}$ and $V_{ts}$ with $V$ the
Cabibbo-Kobayashi-Maskawa matrix) modifies the neutral
current top couplings, without inducing any tree-level correction in
the bottom ones. 
As we will also see, because EWPD is only
sensitive to ${\cal O}_{\phi q}^{(b)}\equiv ({\cal O}_{\phi q}^{(1)}+{\cal O}_{\phi q}^{(3)})_{bb}$
at the tree level, NP effects coming from ${\cal O}_{qq}^{(8)}$ cannot be
constrained by the data. Similarly, EWPD is not sensitive to
$({\cal O}_{\phi u}^{(1)})_{tt}$, 
so ${\cal O}_{uu}^{(1)}$ cannot be constrained if it only involves the third
generation. 

As we mentioned above, we compute the predictions for physical observables
consistently with the approximation of a dimension-six effective
Lagrangian, including only the interference between the SM amplitudes
and the NP effects (i.e. terms linear in $1/\Lambda^2$). We
also use the leading logarithmic approximation for the solution of the
RGE
\be
\frac{d C_i}{d \log{\mu}}=\frac{1}{16\pi^2}\gamma_i^j C_j \Longrightarrow C_i\left(\mu\right)\approx \left(\delta_{i}^{j}+\frac{1}{16\pi^2}\gamma_i^j\left(\Lambda\right)\log{\frac{\mu}{\Lambda}}\right)C_j\left(\Lambda\right),
\label{LlogRGE}
\ee
where we have defined the dimension-six coefficients $C_i\equiv c_i/\Lambda^2$. 
We include in the anomalous dimensions $\gamma_i^j$ the full
dependence on the SM gauge couplings~\cite{Alonso:2013hga} and the
leading contributions from the Yukawa
interactions~\cite{Jenkins:2013wua}, i.e. $Y_e,Y_d\approx 0$,  
and $Y_u\approx\mt{diag}\left(0,0,y_t\right)$. 
The dependence on the Higgs self-coupling~\cite{Jenkins:2013zja} is
irrelevant for our analysis.
Note that, for the contributions to the EWPD that only arise from the RGE, 
within the approximation in Eq.~(\ref{LlogRGE}) 
the physical predictions depend always on $C_i\log{\mu/\Lambda}$.
Finite one-loop contributions beyond the logarithmically enhanced
terms included in the RGE can in some cases be
relevant~\cite{Hartmann:2015oia}, and therefore modify the previous
dependence. Whenever these finite terms are available, we include them in our
analysis. We explicitly comment on these cases below. 


\section{Loop constraints on new top interactions}
\label{constraints:sec}

In this section we present the constraints that EWPD impose on the
operators involving the top quark in (\ref{TopDim6}) due to
RGE effects. We report these bounds in Table~\ref{table:EWPDbounds}
assuming that only one operator is generated at the ultraviolet
scale, $\Lambda$, at a time.\footnote{For completeness, we include here the 
95$\%$ probability interval for the operator coefficient $c_{\phi \Box}$:
\be
\frac{c_{\phi \Box}}{\Lambda^2}\log{\frac{M_Z}{\Lambda}}\in\left[-4.63, 0.65\right]\units{TeV}^{-2}~~\left(c_{\phi \Box}\in\left[-0.27,1.93\right]\mbox{ for }\Lambda=1\units{TeV} \right).
\ee
For negative values of the coefficients, the corresponding bound is then somewhat better than
 the one obtained from Higgs observables~\cite{deBlas:2014ula}.}
We present the results from two
different types of fits. The limits on $C_i \log{M_Z/\Lambda}$ (and $c_i$, for $\Lambda=1\units{TeV}$) are obtained
assuming that the dimension-six coefficients can have either sign. On the other hand,
the bounds on the NP scale $\Lambda$ are derived from a fit with the extra
assumption that the $c_i$ have a definite sign, and then setting this to some illustrative
values, $c_i=\pm 1$.
As we mentioned above, the precision of EWPD overcomes the loop
suppression of the RGE effects and allows to constrain 
most of the interactions at the few percent
level for $\Lambda=1\units{TeV}$, or alternatively pushes the scale
of NP in the top sector to a few TeV for $c_i=\pm1$, thus
fully justifying the use of an effective Lagrangian
description. (Note that, even for the weakest bounds, the
  NP scale is always pushed significantly above the $Z$ mass, 
where the EWPD are measured.)
The leading-log approximation used is also justified provided the value of
$\Lambda$ is not too large, so that
$\left|\frac{\alpha_i}{4\pi}\log{\frac{M_Z}{\Lambda}}\right|\ll 1$, with
$\alpha_i$ the relevant SM parameter. In the rest of this section
we discuss the origin of the constraints on the different operators.

\begin{table}[p]
\begin{center}
{
\begin{tabular}{ l c c c c c c }
\ctoprule
\crowcolor {\bf Operator}&&\multicolumn{2}{c}{\graycell \bf 95$\%$
  prob. interval} &&\multicolumn{2}{c}{\graycell \bf 95$\%$ prob. lower bound}\\
\crowcolor { }&&$\bfmath{\frac{c_i}{\Lambda^2}\log{\frac{M_Z}{\Lambda}}}$&$\bfmath{c_i}$&&\multicolumn{2}{c}{\graycell$\bfmath{\Lambda~\left[\mt{TeV}\right]}$}\\
\crowcolor { }&&$ $&$ $&&$ $&$ $\\[-0.2cm]
\crowcolor { }&&$\bfmath{\left[\mt{TeV}^{-2}\right]}$&$\bfmath{(\Lambda=1\units{TeV})}$&&$\bfmath{(c_i=+1)}$&$\bfmath{(c_i=-1)}$\\
\cmrule
$({\cal O}_{lq}^{(1)})_{eett}$&&$\left[-0.15,0.38\right]$&$\left[-0.16,0.06\right]$&&$4.4$&$3.2$\\
$({\cal O}_{lq}^{(3)})_{eett}$&&$\left[-0.26,0.36\right]$&$\left[-0.15,0.11\right]$&&$3.7$&$3.3$\\
$({\cal O}_{eu})_{eett}$&&$\left[-0.21,0.44\right]$&$\left[-0.18,0.09\right]$&&$3.8$&$2.9$\\
$({\cal O}_{lu})_{eett}$&&$\left[-0.40,0.16\right]$&$\left[-0.07,0.17\right]$&&$3.1$&$4.3$\\
$({\cal O}_{qe})_{ttee}$&&$\left[-0.42,0.20\right]$&$\left[-0.08,0.18\right]$&&$3$&$3.9$\\
\cmrule
$({\cal O}_{lq}^{(1)})_{\mu\mu tt}$&&$\left[-0.91,0.25\right]$&$\left[-0.11,0.38\right]$&&$1.9$&$2.9$\\
$({\cal O}_{lq}^{(3)})_{\mu\mu tt}$&&$\left[-0.04,0.54\right]$&$\left[-0.22,0.02\right]$&&$4.8$&$2.6$\\
$({\cal O}_{eu})_{\mu\mu tt}$&&$\left[-1.29,0.22\right]$&$\left[-0.09,0.54\right]$&&$1.5$&$2.6$\\
$({\cal O}_{lu})_{\mu\mu tt}$&&$\left[-0.26,0.95\right]$&$\left[-0.40,0.11\right]$&&$2.8$&$1.9$\\
$({\cal O}_{qe})_{tt \mu\mu}$&&$\left[-0.22,1.24\right]$&$\left[-0.52,0.09\right]$&&$2.7$&$1.6$\\
\cmrule
$({\cal O}_{lq}^{(1)})_{\tau\tau tt}$&&$\left[-0.52,0.96\right]$&$\left[-0.40,0.22\right]$&&$2.3$&$1.8$\\
$({\cal O}_{lq}^{(3)})_{\tau\tau tt}$&&$\left[-0.86,0.69\right]$&$\left[-0.29,0.36\right]$&&$1.9$&$2.1$\\
$({\cal O}_{eu})_{\tau\tau tt}$&&$\left[-0.58,1.18\right]$&$\left[-0.49,0.24\right]$&&$2.1$&$1.6$\\
$({\cal O}_{lu})_{\tau\tau tt}$&&$\left[-1.01,0.54\right]$&$\left[-0.23,0.42\right]$&&$1.8$&$2.2$\\
$({\cal O}_{qe})_{tt \tau\tau}$&&$\left[-1.14,0.56\right]$&$\left[-0.23,0.48\right]$&&$1.7$&$2.2$\\
\cmrule
$({\cal O}_{lq}^{(1)})_{\ell\ell tt}$&&$\left[-0.16,0.26\right]$&$\left[-0.11,0.07\right]$&&$4.7$&$3.9$\\
$({\cal O}_{lq}^{(3)})_{\ell\ell tt}$&&$\left[-0.07,0.29\right]$&$\left[-0.12,0.03\right]$&&$5.9$&$3.8$\\
$({\cal O}_{eu})_{\ell\ell tt}$&&$\left[-0.24,0.33\right]$&$\left[-0.14,0.10\right]$&&$3.8$&$3.4$\\
$({\cal O}_{lu})_{\ell\ell tt}$&&$\left[-0.27,0.17\right]$&$\left[-0.07,0.11\right]$&&$3.8$&$4.6$\\
$({\cal O}_{qe})_{ tt\ell\ell}$&&$\left[-0.32,0.23\right]$&$\left[-0.10,0.13\right]$&&$3.4$&$3.9$\\
\cmrule
$({\cal O}_{qq}^{(1)})_{tttt}$&&$\left[-0.55,1.38\right]$&$\left[-0.58,0.23\right]$&&$2.1$&$1.5$\\
$({\cal O}_{ud}^{(1)})_{ttbb}$&&$\left[0.25,10.9\right]$&$\left[-4.6,-0.10\right]$&&$0.89$&$0.37$\\
$({\cal O}_{qu}^{(1)})_{tttt}$&&$\left[-1.47,0.59\right]$&$\left[-0.25,0.62\right]$&&$1.4$&$2$\\
$({\cal O}_{qd}^{(1)})_{ttbb}$&&$\left[-9.7,-0.07\right]$&$\left[0.03,4.06\right]$&&$0.41$&$0.95$\\
\cmrule
$({\cal O}_{uB})_{tt}$&&$\left[-0.35,0.10\right]$&$\left[-0.04,0.15\right]$&&$3.4$&$5.1$\\
$({\cal O}_{uW})_{tt}$&&$\left[-0.39,0.11\right]$&$\left[-0.05,0.17\right]$&&$3.2$&$4.7$\\
\cbottomrule  
\end{tabular}}
\caption{ 
EWPD bounds on top interactions, assuming one operator at a time at the scale $\Lambda$. 
The bounds on the NP scale $\Lambda$ are obtained
  from two independent types of fits, assuming 
  a definite sign for the coefficients $c_i$.
The results for the operators $\left({\cal O}_i\right)_{\ell\ell tt,
    tt\ell\ell}$ are obtained assuming lepton universality in the
  interactions. The bounds for $({\cal O}_{qq}^{(8)})_{tttt}$
are too weak and have been omitted, while the operator coefficient for $({\cal O}_{uu}^{(1)})_{tttt}$ 
cannot be constrained within our approximations (see text for details).
 }
\label{table:EWPDbounds}
\end{center}
\end{table}

The relatively strong constraints on $\ell^+ \ell^- t\bar{t}$
interactions can be understood from the fact that all those
interactions contribute to the running of $c_{\phi l}^{(1)}$, $c_{\phi
  l}^{(3)}$ or $c_{\phi e}^{(1)}$. The corresponding operators provide direct corrections to
the neutral current couplings of the charged leptons, and are bounded at the
per mile level (see~\cite{Carpentier:2010ue} for an earlier partial
analysis). Note that the bounds for some of these interactions,
e.g. ${\cal O}_{lq}^{(1)}$ and ${\cal O}_{lu}$, are almost identical (up to
a sign). This correlation follows directly from the RGE for the
leptonic interactions in (\ref{EWPDdim6}). In particular, 
\bea
\frac{d (C_{\phi l}^{(1)})_{ij}}{d
  \log{\mu}}&=&\frac{N_c}{8\pi^2}\left\{\left(Y_u^\dagger
Y_u\right)_{lk}\left(C_{lq}^{(1)}\right)_{ijkl}-\left(Y_u
Y_u^\dagger\right)_{lk}\left(C_{lu}\right)_{ijkl}\right\}+\ldots,\nonumber
\\ 
\frac{d (C_{\phi e}^{(1)})_{ij}}{d \log{\mu}}&=&\frac{N_c}{8\pi^2}\left\{\left(Y_u^\dagger Y_u\right)_{lk}\left(C_{qe}\right)_{klij}-\left(Y_u Y_u^\dagger\right)_{lk}\left(C_{eu}\right)_{ijkl}\right\}+\ldots~.
\label{RGEcorrleptonic}
\eea
Thus, only the combinations of operators appearing in 
\refeq{RGEcorrleptonic} can be constrained by EWPD, up to corrections in the
RGE induced by the gauge interactions. 

Constraints on four-quark interactions involving only the third family are
dominated by the contributions they generate to the $Zb\bar{b}$
couplings, 
via the operators ${\cal O}_{\phi q}^{(1),(3)}$ and ${\cal O}_{\phi d}^{(1)}$, and
are therefore somewhat weaker than the leptonic ones. Limits on
$c_{qq}^{(1)}$ and $c_{qu}^{(1)}$ arise from the bounds on the left-handed
bottom couplings, and are significantly stronger than those of
$c_{ud}^{(1)}$ and $c_{qd}^{(1)}$, which contribute to the $Zb_R
\overline{b_R}$ interactions. In particular, the strong preference for a positive
(negative) value of $(c_{ud}^{(1)})_{ttbb}$ ($(c_{qd}^{(1)})_{ttbb}$)
follows from the corresponding preference for a large correction to
the right-handed bottom coupling, $\delta g_R^b=-\frac 12(c_{\phi
  d}^{(1)})_{bb}\frac{v^2}{\Lambda^2}$, with $v\approx 246$ GeV, to alleviate the
$-2.5$-$\sigma$ deviation in the bottom forward-backward asymmetry at
the $Z$-pole.
Again, some of the bounds on these four-quark operators 
can be easily correlated from
the corresponding contributions to the running for the quark
interactions in (\ref{EWPDdim6}), 
\bea
\frac{d (C_{\phi q}^{(1)}+C_{\phi q}^{(3)})_{ij}}{d \log{\mu}}\!\!\!\!&=&\!\!\!\!\frac{N_c}{16\pi^2}\!\left\{\!\left(Y_u^\dagger Y_u\right)_{lk}\!\left(\!\left(C_{qq}^{(1)}\right)_{ijkl}+\left(C_{qq}^{(1)}\right)_{klij}\right)\!-\!2\left(Y_u Y_u^\dagger\right)_{lk}\left(C_{qu}^{(1)}\right)_{ijkl}\right\}\!+\!\ldots,\nn
\frac{d (C_{\phi d}^{(1)})_{ij}}{d \log{\mu}}\!\!\!\!&=&\!\!\!\!\frac{N_c}{8\pi^2}\left(Y_u^\dagger Y_u\right)_{lk}\left(\!\left(C_{qd}^{(1)}\right)_{klij}-\left(C_{ud}^{(1)}\right)_{klij}\!\right)+\ldots~,
\label{RGEcorrquark}
\eea
that determine which combinations of operators can be constrained by
EWPD. In the first line of \refeq{RGEcorrquark} there is no contribution 
from ${\cal O}_{qq}^{(8)}$, because the corresponding corrections
to the running of $C_{\phi q}^{(1)}$ and $C_{\phi q}^{(3)}$ cancel each
other. 
There is a suppressed contribution to the running of $C_{\phi
  q}^{(1)}+C_{\phi 
  q}^{(3)}$ from $C_{qq}^{(8)}$ proportional to the
electroweak gauge couplings, which results in much weaker
constraints. This explains the absence of a 
bound on $C_{qq}^{(8)}$ in Table~\ref{table:EWPDbounds}.
Finally, the coefficient $(C_{uu}^{(1)})_{tttt}$ only
contributes to the $Z t_R\overline{t_R}$ couplings through the RGE 
for $(C_{\phi u}^{(1)})_{tt}$, and therefore cannot be bounded by
EWPD at the order we are working.

Four-quark operators involving two quarks of the third generation 
and two of either the first or second generations
contribute, through RGE, to operators that modify the electroweak couplings of the
quarks in the first two generations. These have been
measured with worse precision than those of the charged leptons or
bottom quark. Hence, the corresponding bounds
are much weaker and not reported here. If one
assumes universality among the three families then the bounds are
still mostly dominated by the operators involving only third generation
quarks. The exception is the case of the operators ${\cal O}^{(1)}_{qd,ud}$, for
which there is a tension between the required contribution to
$\delta g_R^b$ and the corresponding one for the first two
generations. This tension results in significantly improved bounds,
reducing the  
size of the corresponding 95$\%$ probability intervals by a factor of two, e.g.
$(C_{qd}^{(1)})_{ttqq}\log\frac{M_Z}{\Lambda} \in[-4.09,0.65]$ TeV$^{-2}$.

The limits on the electroweak top dipole interactions, $({\cal O}_{uB})_{tt}$ and
$({\cal O}_{uW})_{tt}$, come exclusively from their contributions to the running of
$c_{WB}$ ($c_{WB}/\Lambda^2 \in
\left[-0.009,0.003\right]\units{TeV}^{-2}$ at 95\% probability), which is
related to the $S$ parameter~\cite{Peskin:1991sw}.
Hence, only
  the approximate combination 
$g_2 (C_{uB})_{tt} + 2 g_1(\frac{1}{6}+\frac{2}{3})(C_{uW})_{tt}$
  (where the $1/6$ and $2/3$ factors 
are the $q_L$ and $u_R$ hypercharges, respectively),
  which enters in the RGE for $c_{WB}$, can be constrained.

Finally, we have not included in Table~\ref{table:EWPDbounds} the
constraints on the operators that induce direct corrections to the
top electroweak couplings, 
\be
\delta g_L^t=-\frac 12\left(V\left(c_{\phi q}^{(1)}-c_{\phi q}^{(3)}\right)V^\dagger\right)_{tt}\frac{v^2}{\Lambda^2}=-c_{\phi q}^{(t)}\frac{v^2}{\Lambda^2},~~~~~~~\delta g_R^t=-\frac 12\left(c_{\phi u}^{(1)}\right)_{tt}\frac{v^2}{\Lambda^2}.
\label{dgt}
\ee
Note that,
  to dimension six, the effects on the left-handed sector are also correlated
  with the direct corrections of the charged current couplings,
  $\delta V_{tb}=(Vc_{\phi q}^{(3)})_{tb}v^2/\Lambda^2$. (We work in
a flavor basis in which the SM Yukawa couplings for the down sector
are diagonal.)   The constraints on the combinations in \refeq{dgt}
follow from the one-loop contributions to the $T$ parameter and
corrections to the $Zb\bar{b}$ vertices. These corrections contain
logarithmically enhanced terms that can be read from the RGE of the
operators ${\cal O}_{\phi D}$  
(equivalent to the $T$ parameter in our basis), 
${\cal O}_{\phi q}^{(b)}\equiv ({\cal O}_{\phi q}^{(1)}+{\cal O}_{\phi q}^{(3)})_{bb}$ and
$({\cal O}_{\phi d}^{(1)})_{bb}$. We have also included finite (not
proportional to logarithms) one-loop
effects by integrating out the top quark with the anomalous couplings
defined in
\refeq{dgt}~\cite{Pomarol:2008bh,Anastasiou:2009rv}. In particular, the finite
contribution to the $T$ parameter is given by
\be
\alpha \Delta T=\frac{N_c}{16\pi^2}y_t^2\mt{Re}\left\{(V\alpha_{\phi q}^{(3)})_{tb} V_{tb}^*\right\}\frac{v^2}{\Lambda^2}.
\ee
Because of these 
finite terms, the $\chi^2$ 
depends on both $C_i$ and $C_i \log{\frac{M_Z}{\Lambda}}$, so we vary $C_i$ and $\Lambda$ independently in our fits. We impose $\Lambda\geq 1\units{TeV}$, to avoid regions where the effective Lagrangian description may break down.
In the bounds below, when no mention to $\Lambda$ is made, we take the most conservative bound that is reached for $\Lambda=1$ TeV.

Considering only one of the combinations in \refeq{dgt} at a time we obtain the following 95\% probability interval for $\delta g_L^t/g_L^{t~\!\mbox{\tiny SM}}$,
\be
\!\!\!\!\!\!\!\!\!\!\frac{\delta g_L^t}{g_L^{t~\!\mbox{\tiny SM}}}\in \left[-0.016,0.002\right]~~~\left(c_{\phi q}^{(t)}\in \left[-0.01,0.09\right]\right),
\label{Zttbounds1DL}
\ee
while for $\delta g_R^t/g_R^{t~\!\mbox{\tiny SM}}$ we get
\be
\!\frac{\delta g_R^t}{g_R^{t~\!\mbox{\tiny SM}}}\in \left[-0.017,0.002\right]~~~\left((c_{\phi u}^{(1)})_{tt}\in \left[-0.08,0.01\right]\right),
\label{Zttbounds1DR}
\ee
where $g_L^{t~\!\mbox{\tiny SM}}=\frac 12 -\frac 23 \sin^2{\theta_W}$ and $g_R^{t~\!\mbox{\tiny SM}}= -\frac 23 \sin^2{\theta_W}$, with $\theta_W$ the weak angle.
When we consider dimension-six effects correcting both electroweak top couplings at the same time, the 95$\%$ probability bounds change to:
\be
\begin{split}
~~\frac{\delta g_L^t}{g_L^{t~\!\mbox{\tiny SM}}}\in \left[-0.048,0.089\right],&~~~~~~~~\frac{\delta g_R^t}{g_R^{t~\!\mbox{\tiny SM}}}\in \left[-0.102,0.044\right].\\
\!\!\!\left(c_{\phi q}^{(t)}\in \left[-0.52,0.28\right],\right.&~~~~~~\!\left.(c_{\phi u}^{(1)})_{tt}\in \left[-0.50,0.21\right].\right)
\end{split}
\label{Zttbounds2D}
\ee
\begin{figure}[t]
\begin{center}
\input{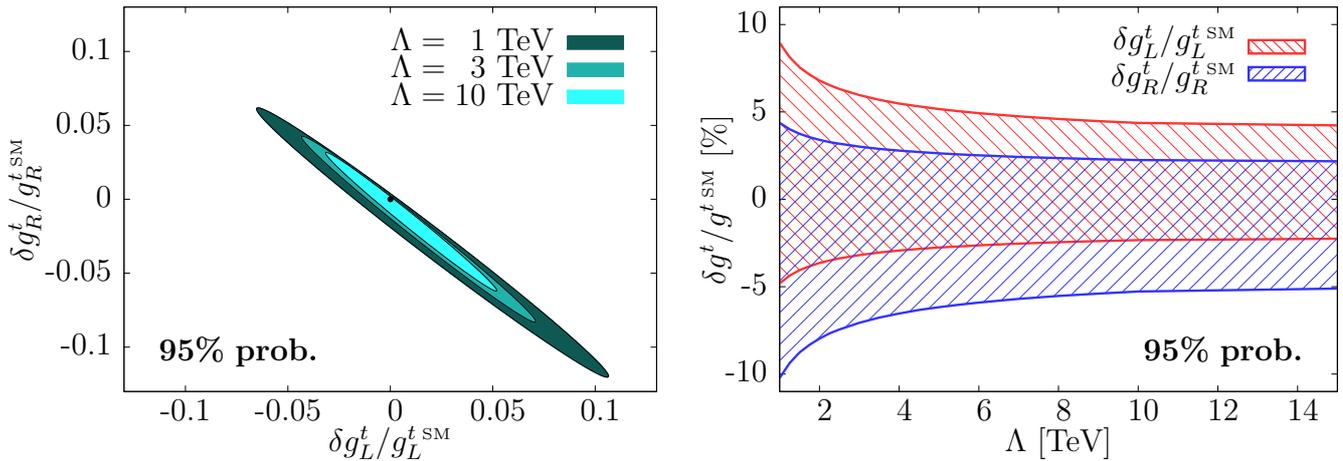}
\end{center}
\caption{{\it Left)} 95$\%$ probability regions in the $\frac{\delta g_L^t}{g_L^{t~\!\mbox{\tiny SM}}}-\frac{\delta g_R^t}{g_R^{t~\!\mbox{\tiny SM}}}$ plane for $\Lambda=$1, 3 and 10 TeV. {\it Right)} Boundaries of the 95$\%$ probability intervals for $\frac{\delta g_{L,R}^t}{g_{L,R}^{t~\!\mbox{\tiny SM}}}$ as a function of $\Lambda$.}
\label{fig:gLtgRt}
\end{figure}
The weaker bounds compared with Eqs.~(\ref{Zttbounds1DL}) and (\ref{Zttbounds1DR}) follow from a strong correlation ($\approx\!-99\%$) between these two couplings, as can be seen in
Figure~\ref{fig:gLtgRt} left, which can be understood from the leading
logarithmic contributions. Indeed, neglecting the finite contributions,
there is a large correlation between the corresponding dimension-six
operators, which comes from the RGE for $C_{\phi D}$ (whose limit
dominates the constraints in Eqs.~(\ref{Zttbounds1DL}) and (\ref{Zttbounds1DR}) via the $T$ parameter),  
\be
\frac{d C_{\phi D}}{d \log{\mu}}=\frac{N_c}{2\pi^2}\left\{\left(C_{\phi q}^{(1)}\right)_{ij}\left(Y_u^\dagger Y_u\right)_{ji}- \left(Y_u^\dagger Y_u\right)_{ij}\left(C_{\phi u}^{(1)}\right)_{ji}\right\}+\ldots~.
\label{RGET}
\ee
At this level there is an approximate flat direction for 
$\delta g_L^t/g_L^{t~\!\mbox{\tiny SM}}=-\delta
g_R^t/g_R^{t~\!\mbox{\tiny SM}}$ (we have used 
$g_L^{t~\!\mbox{\tiny SM}}\approx -2 g_R^{t~\!\mbox{\tiny SM}}$ and
$c_{\phi q}^{(3)} = - c_{\phi q}^{(1)}$), which
is also reflected in the equality of the bounds in Eqs.~(\ref{Zttbounds1DL}) and (\ref{Zttbounds1DR}).
This flat direction is lifted however by the logarithmic contributions to the $Zb_L \overline{b_L}$
vertex and, to a less extent, the contributions in the running from gauge 
interactions. The global factors in
front of the finite terms turn out to be smaller than the
ones in the logarithmic terms, so the effects from the former are not very important
for values of $\Lambda$ consistent with the effective Lagrangian description.
In Figure~\ref{fig:gLtgRt} right, we show how the bounds
obtained in \refeq{Zttbounds2D} for $\Lambda=1\units{TeV}$ evolve as we 
increase the value of the scale of NP.

The results in \refeq{Zttbounds2D} imply quite strong bounds
on deviations with the respect to the SM electroweak theory, at the few 
percent level. 
One may wonder though about the robustness of these bounds when interpreted
within particular models. For instance,
as explained above, the effects in the $T$
parameter have a strong impact in these bounds. Now, the $T$ parameter
is known to have a strong correlation with the $S$
parameter (we obtain an $89\%$ correlation), and many
SM extensions in which 
the NP interacts mainly with the third generation
come also usually accompanied by relatively large positive
contributions to $S$.
Still, considering a positive value for $S=0.2$ (about the $95\%$
probability limit 
obtained from the $S$-$T$ fit) does not have a dramatic impact on the
bounds in \refeq{Zttbounds2D}, due to the constraints
on the logarithmic contributions to $Zb_L \overline{b_L}$.  
We thus obtain only a moderate shift in the bounds for the right-handed
couplings,
\be
\frac{\delta g_L^t}{g_L^{t~\!\mbox{\tiny SM}}}\in \left[-0.050,0.088\right],~~~~~~\frac{\delta g_R^t}{g_R^{t~\!\mbox{\tiny SM}}}\in \left[-0.123,0.023\right]~~~(S=0.2),
\label{Zttbounds2DS02}
\ee
but we can still conclude that NP contributions to $Zt\bar{t}$
couplings beyond 10$\%$ are disfavored by the data. 


\section{Conclusions}
\label{conclusions:sec}

Among the possible NP deformations of the SM there are several on which no
direct experimental information can be extracted from present or past
experiments. Indirect constraints can however be obtained in some
cases from their loop contribution to
the very precisely measured EWPD. Using a model-independent effective
Lagrangian approach, we have explored in this article the potential that
EWPD have to put constraints on dimension-six effective operators on
which no precise direct information is currently available. We have
shown that, despite the loop suppression, the leading contributions
proportional to the top Yukawa coupling are large enough to place
significant bounds on several interactions involving the top quark.

Our results, reported in Table~\ref{table:EWPDbounds}, show that, for
NP in the TeV region, EWPD can constrain the dimension-six
operator coefficients for a large set of interactions involving the
top quark to $O(0.1)$ values. 
These results are obtained using the logarithmically enhanced one-loop
contributions, as given by the RGE, together with the finite terms in
the most significant cases. 
Barring accidental cancellations, similar bounds will apply for
all the interactions we have considered, even if the missing finite
terms give a comparable effect to the logarithmically enhanced ones. 

The bounds presented here have been computed assuming only one operator 
at a time at the scale of NP. There are several 
scenarios with 
new scalars, quarks or vector bosons for which such operators can be
generated alone, or whose effects are not correlated with the
contributions to other dimension-six
interactions~\cite{deBlas:2014mba,delAguila:2000rc,delAguila:2010mx}. 
Nevertheless, we have discussed the origin of the leading constraints
and described the most relevant approximate flat directions so that
bounds on more complicated models can be estimated.
As an example, let us discuss the case of lepton-top four-fermion
interactions. 
For NP at the TeV scale we are able to place
$O(0.1)$ bounds on the coefficients of contact interactions resulting
from the products of vector currents involving two electron and two top
fields, with all possible chiralities. As we have mentioned, 
the effects of some of these operators are
correlated in the RGE that enter in the EWPD so, in practice, we can
only constrain three combinations of these $e^+ e^- t\bar{t}$ operators. Removing the two redundant interactions from
a simultaneous global fit to all the operators we obtain the following
limits (for $\Lambda=1\units{TeV}$): 
\begin{eqnarray}\nonumber
(c_{lq}^{(1)}-c_{lu})_{eett}&\in&\left[-0.21,0.08\right],\\
(c_{lq}^{(3)})_{eett}&\in&\left[-0.31,0.06\right],\nonumber\\
(c_{qe}-c_{eu})_{eett}&\in&\left[-0.10,0.29\right].
\end{eqnarray}
In particular, these $e^+ e^- t\bar{t}$ contact interactions can be
tested in future 
colliders~\cite{Grzadkowski:1997cj}, like the ILC~\cite{Baer:2013cma}
or the FCC-ee~\cite{Gomez-Ceballos:2013zzn}, and are therefore
relevant to guide NP 
searches in these facilities. The same applies for our analysis of the $Zt\bar{t}$ couplings, which could be directly measured at this kind of experiments with great precision (around 1$\%$ for the ILC with 500 fb$^{-1}$~\cite{Devetak:2010na}). Finally, while less precise, our
limits on four-top interactions are comparable (and in some cases
significantly better) than 
the latest ones obtained by the LHC~\cite{Aad:2015kqa}.

Higher-dimensional operators involving the top quark are not only
among the least constrained operators that contribute the most to
EWPD, they are also among the best motivated ones in models that
attempt to solve the hierarchy problem. 
In many of these models, a
sizeable contribution to the $S$ parameter is generated at the NP
 scale. We have considered that possibility, see
Eq.~(\ref{Zttbounds2DS02}), and shown that while inducing a shift in
the allowed values of the coefficients of the higher-dimensional
operators involving the top, these can still be constrained to a 
similar level of accuracy.
This also illustrates 
the fact that, being indirect bounds, the limits we have computed are
sensitive to assumptions about operators that can give tree-level
contributions to EWPD. In the presence of such operators the
quantitative results may change but the qualitative fact that
one-loop contributions can place the most stringent bounds on currently
untested operators still holds. These bounds are therefore a crucial
piece of information, both for model building purposes and as a guide
for future experimental searches.\\


\textbf{Note added:} While this manuscript was being prepared for submission, ref.~\cite{Buckley:2015nca} appeared in the arXiv. In that work, a global fit to dimension-six effective operators involving the top quark is performed using top production data from the LHC and the Tevatron. Our approach is complementary to theirs, in the sense that our analysis is sensitive to a different class of operators. In the few cases in which there is overlap, our results provide more stringent bounds on the coefficients of the operators.

\newpage
\section*{Acknowledgements}

We thank L. Silvestrini for useful discussions.
The work of JB has been supported by the European Research Council
under the European Union's Seventh Framework Programme
(FP~\!\!/~\!\!2007-2013)~\!/~\!ERC Grant Agreement n. 279972. 
The work of JS has been partially supported by the European Commission
(PITN-GA-2012-316704 HIGGSTOOLS), by MINECO under grants
(FPA2010-17915 and FPA2013-47836-C3-2-P) and by Junta de
Andaluc\'{\i}a grants FQM 101 and FQM 6552.
JS thanks the Pauli Center Visitor Program for financial support. 




\begin{thebibliography}{99}

\bibitem{deBlas:2013qqa}
  J.~de Blas, M.~Chala and J.~Santiago,
  Phys.\ Rev.\ D {\bf 88} (2013) 095011
  [arXiv:1307.5068 [hep-ph]].

\bibitem{eff:top}
  J.~A.~Aguilar-Saavedra,
  Nucl.\ Phys.\ B {\bf 821} (2009) 215
  [arXiv:0904.2387 [hep-ph]];
  Nucl.\ Phys.\ B {\bf 821} (2009) 215
  [arXiv:0904.2387 [hep-ph]];
  Nucl.\ Phys.\ B {\bf 843} (2011) 638
   [Nucl.\ Phys.\ B {\bf 851} (2011) 443]
  [arXiv:1008.3562 [hep-ph]];
  C.~Zhang and S.~Willenbrock,
  Phys.\ Rev.\ D {\bf 83} (2011) 034006
  [arXiv:1008.3869 [hep-ph]];
  C.~Degrande, J.~M.~Gerard, C.~Grojean, F.~Maltoni and G.~Servant,
  JHEP {\bf 1103} (2011) 125
  [arXiv:1010.6304 [hep-ph]];
  J.~A.~Aguilar-Saavedra and M.~Perez-Victoria,
  JHEP {\bf 1105} (2011) 034
  [arXiv:1103.2765 [hep-ph]];
  C.~Degrande, J.~M.~Gerard, C.~Grojean, F.~Maltoni and G.~Servant,
  Phys.\ Lett.\ B {\bf 703} (2011) 306
  [arXiv:1104.1798 [hep-ph]];
  JHEP {\bf 1207} (2012) 036
   [JHEP {\bf 1303} (2013) 032]
  [arXiv:1205.1065 [hep-ph]];
  M.~Fabbrichesi, M.~Pinamonti and A.~Tonero,
  Phys.\ Rev.\ D {\bf 89} (2014) 7,  074028
  [arXiv:1307.5750 [hep-ph]];
  Eur.\ Phys.\ J.\ C {\bf 74} (2014) 12,  3193
  [arXiv:1406.5393 [hep-ph]];
  Q.~H.~Cao, B.~Yan, J.~H.~Yu and C.~Zhang,
  arXiv:1504.03785 [hep-ph].

\bibitem{eff:top:nlo}
  C.~Zhang,
  Phys.\ Rev.\ D {\bf 90} (2014) 1,  014008
  [arXiv:1404.1264 [hep-ph]];
  C.~Zhang,
  J.\ Phys.\ Conf.\ Ser.\  {\bf 556} (2014) 1,  012030
  [arXiv:1410.2825 [hep-ph]];
  G.~Durieux, F.~Maltoni and C.~Zhang,
  Phys.\ Rev.\ D {\bf 91} (2015) 7,  074017
  [arXiv:1412.7166 [hep-ph]].

\bibitem{De Rujula:1991se}
  A.~De Rujula, M.~B.~Gavela, P.~Hernandez and E.~Masso,
  Nucl.\ Phys.\ B {\bf 384} (1992) 3;
%
  K.~Hagiwara, S.~Ishihara, R.~Szalapski and D.~Zeppenfeld,
  Phys.\ Lett.\ B {\bf 283} (1992) 353;
%
  Phys.\ Rev.\ D {\bf 48} (1993) 2182;
%
  S.~Alam, S.~Dawson and R.~Szalapski,
  Phys.\ Rev.\ D {\bf 57} (1998) 1577
  [hep-ph/9706542];
%
  H.~Mebane, N.~Greiner, C.~Zhang and S.~Willenbrock,
  Phys.\ Rev.\ D {\bf 88} (2013) 1,  015028
  [arXiv:1306.3380 [hep-ph]].


\bibitem{Jenkins:2013zja}
  E.~E.~Jenkins, A.~V.~Manohar and M.~Trott,
  JHEP {\bf 1310} (2013) 087
  [arXiv:1308.2627 [hep-ph]].
  
\bibitem{Jenkins:2013wua}
  E.~E.~Jenkins, A.~V.~Manohar and M.~Trott,
  JHEP {\bf 1401} (2014) 035
  [arXiv:1310.4838 [hep-ph]].
  
\bibitem{Alonso:2013hga}
  R.~Alonso, E.~E.~Jenkins, A.~V.~Manohar and M.~Trott,
  JHEP {\bf 1404} (2014) 159
  [arXiv:1312.2014 [hep-ph]].

\bibitem{Grojean:2013kd}
  C.~Grojean, E.~E.~Jenkins, A.~V.~Manohar and M.~Trott,
  JHEP {\bf 1304} (2013) 016
  [arXiv:1301.2588 [hep-ph]];
%
  J.~Elias-Mir\'o, J.~R.~Espinosa, E.~Masso and A.~Pomarol,
  JHEP {\bf 1308} (2013) 033
  [arXiv:1302.5661 [hep-ph]];
  JHEP {\bf 1311} (2013) 066
  [arXiv:1308.1879 [hep-ph]];
  J.~Elias-Mir\'o, C.~Grojean, R.~S.~Gupta and D.~Marzocca,
  JHEP {\bf 1405} (2014) 019
  [arXiv:1312.2928 [hep-ph]].


\bibitem{Grzadkowski:2010es}
  B.~Grzadkowski, M.~Iskrzynski, M.~Misiak and J.~Rosiek,
  JHEP {\bf 1010} (2010) 085
  [arXiv:1008.4884 [hep-ph]].


\bibitem{Aad:2015zhl}
  G.~Aad {\it et al.}  [ATLAS and CMS Collaborations],
  Phys.\ Rev.\ Lett.\  {\bf 114} (2015) 191803
  [arXiv:1503.07589 [hep-ex]];
%
  [ATLAS and CDF and CMS and D0 Collaborations],
  arXiv:1403.4427 [hep-ex];
%
  K.~A.~Olive {\it et al.}  [Particle Data Group Collaboration],
  Chin.\ Phys.\ C {\bf 38} (2014) 090001;
%
  M.~Davier, A.~Hoecker, B.~Malaescu and Z.~Zhang,
  Eur.\ Phys.\ J.\ C {\bf 71} (2011) 1515
   [Eur.\ Phys.\ J.\ C {\bf 72} (2012) 1874]
  [arXiv:1010.4180 [hep-ph]];
%
  T.~E.~W.~Group [CDF and D0 Collaborations],
  arXiv:1204.0042 [hep-ex];
%
  LEP Electroweak Working Group [ALEPH and CDF and D0 and DELPHI and L3 and OPAL and SLD and LEP Electroweak Working Group and Tevatron Electroweak Working Group and SLD Electroweak and Heavy Flavour Groups Collaborations],
  arXiv:1012.2367 [hep-ex];
%
  S.~Schael {\it et al.}  [ALEPH and DELPHI and L3 and OPAL and SLD and LEP Electroweak Working Group and SLD Electroweak Group and SLD Heavy Flavour Group Collaborations],
  Phys.\ Rept.\  {\bf 427} (2006) 257
  [hep-ex/0509008].
  

\bibitem{Awramik:2003rn}
  M.~Awramik, M.~Czakon, A.~Freitas and G.~Weiglein,
  Phys.\ Rev.\ D {\bf 69} (2004) 053006
  [hep-ph/0311148];
%
  M.~Awramik, M.~Czakon and A.~Freitas,
  JHEP {\bf 0611} (2006) 048
  [hep-ph/0608099];
%
  M.~Awramik, M.~Czakon, A.~Freitas and B.~A.~Kniehl,
  Nucl.\ Phys.\ B {\bf 813} (2009) 174
  [arXiv:0811.1364 [hep-ph]];
%
  A.~Freitas,
  JHEP {\bf 1404} (2014) 070
  [arXiv:1401.2447 [hep-ph]].
  

\bibitem{Berthier:2015oma}
  L.~Berthier and M.~Trott,
  JHEP {\bf 1505} (2015) 024
  [arXiv:1502.02570 [hep-ph]].
  

\bibitem{Han:2004az}
  Z.~Han and W.~Skiba,
  Phys.\ Rev.\ D {\bf 71} (2005) 075009
  [hep-ph/0412166];
%
  F.~del Aguila and J.~de Blas,
  Fortsch.\ Phys.\  {\bf 59} (2011) 1036
  [arXiv:1105.6103 [hep-ph]];
%
  M.~Ciuchini, E.~Franco, S.~Mishima and L.~Silvestrini,
  JHEP {\bf 1308} (2013) 106
  [arXiv:1306.4644 [hep-ph]];
%
  J.~de Blas,
  EPJ Web Conf.\  {\bf 60} (2013) 19008
  [arXiv:1307.6173 [hep-ph]];
  A.~Pomarol and F.~Riva,
  JHEP {\bf 1401} (2014) 151
  [arXiv:1308.2803 [hep-ph]];
%
  J.~Ellis, V.~Sanz and T.~You,
  JHEP {\bf 1503} (2015) 157
  [arXiv:1410.7703 [hep-ph]];
  A.~Efrati, A.~Falkowski and Y.~Soreq,
  arXiv:1503.07872 [hep-ph].

 
\bibitem{deBlas:2014mba}
  J.~de Blas, M.~Chala, M.~Perez-Victoria and J.~Santiago,
  JHEP {\bf 1504} (2015) 078
  [arXiv:1412.8480 [hep-ph]].


\bibitem{UpdateEWfitdim6}
 J.~de Blas et al. In preparation.
 

\bibitem{deBlas:2014ula}
  J.~de Blas, M.~Ciuchini, E.~Franco, D.~Ghosh, S.~Mishima, M.~Pierini, L.~Reina and L.~Silvestrini,
  arXiv:1410.4204 [hep-ph].

\bibitem{Zhang:2012cd}
  C.~Zhang, N.~Greiner and S.~Willenbrock,
  Phys.\ Rev.\ D {\bf 86} (2012) 014024
  [arXiv:1201.6670 [hep-ph]].


\bibitem{Hartmann:2015oia}
  C.~Hartmann and M.~Trott,
  arXiv:1505.02646 [hep-ph];
  M.~Ghezzi, R.~Gomez-Ambrosio, G.~Passarino and S.~Uccirati,
  arXiv:1505.03706 [hep-ph].

\bibitem{Carpentier:2010ue}
  M.~Carpentier and S.~Davidson,
  Eur.\ Phys.\ J.\ C {\bf 70} (2010) 1071
  [arXiv:1008.0280 [hep-ph]].
 
  
\bibitem{Peskin:1991sw}
  M.~E.~Peskin and T.~Takeuchi,
  Phys.\ Rev.\ D {\bf 46} (1992) 381.

\bibitem{Pomarol:2008bh}
  A.~Pomarol and J.~Serra,
  Phys.\ Rev.\ D {\bf 78} (2008) 074026
  [arXiv:0806.3247 [hep-ph]].

\bibitem{Anastasiou:2009rv}
  C.~Anastasiou, E.~Furlan and J.~Santiago,
  Phys.\ Rev.\ D {\bf 79} (2009) 075003
  [arXiv:0901.2117 [hep-ph]].
  
  
\bibitem{delAguila:2000rc}
  F.~del Aguila, M.~Perez-Victoria and J.~Santiago,
  JHEP {\bf 0009} (2000) 011
  [hep-ph/0007316].
  
 
\bibitem{delAguila:2010mx}
  F.~del Aguila, J.~de Blas and M.~Perez-Victoria,
  JHEP {\bf 1009} (2010) 033
  [arXiv:1005.3998 [hep-ph]].
  
\bibitem{Grzadkowski:1997cj}
  B.~Grzadkowski, Z.~Hioki and M.~Szafranski,
  Phys.\ Rev.\ D {\bf 58} (1998) 035002
  [hep-ph/9712357].


\bibitem{Baer:2013cma}
  H.~Baer, T.~Barklow, K.~Fujii, Y.~Gao, A.~Hoang, S.~Kanemura, J.~List and H.~E.~Logan {\it et al.},
  arXiv:1306.6352 [hep-ph].
  

\bibitem{Gomez-Ceballos:2013zzn}
  M.~Bicer {\it et al.}  [TLEP Design Study Working Group Collaboration],
  JHEP {\bf 1401} (2014) 164
  [arXiv:1308.6176 [hep-ex]].

\bibitem{Devetak:2010na}
  E.~Devetak, A.~Nomerotski and M.~Peskin,
  Phys.\ Rev.\ D {\bf 84} (2011) 034029
  [arXiv:1005.1756 [hep-ex]].
  

\bibitem{Aad:2015kqa}
  G.~Aad {\it et al.}  [ATLAS Collaboration],
  arXiv:1505.04306 [hep-ex].
  
\bibitem{Buckley:2015nca}
  A.~Buckley, C.~Englert, J.~Ferrando, D.~J.~Miller, L.~Moore, M.~Russell and C.~D.~White,
  arXiv:1506.08845 [hep-ph].
  
\end{thebibliography}
\end{document}